\documentclass[longauth,letter]{aa}  
\usepackage{epsfig}
\usepackage{natbib}
\usepackage{txfonts}
\usepackage{graphicx}

\def\ms{\hbox{\,m/s}}         
\def\m2s2{\hbox{\,m$^{2}$\,s$^{-2}$}} 
\def\kms{\hbox{\,km/s}}       
\def\Msun{\hbox{$M_{\odot}$}}             
\def\Rsun{\hbox{$R_{\odot}$}}
\def\Mjup{\hbox{$M_{\rm Jup}$}}
\def\Rjup{\hbox{$R_{\rm Jup}$}}

\def\gcmmm{\hbox{\,g\,cm$^{-3}$}}         

\begin{document}

\title{Transiting exoplanets from the CoRoT space mission}
\subtitle{V. CoRoT-Exo-4b: Stellar and planetary parameters 
\thanks{Based on observations obtained with CoRoT, a space project operated by the French Space Agency, CNES, with participation of the Science Programme of ESA, ESTEC/RSSD, Austria, Belgium, Brazil, Germany and Spain; and on observations made with the SOPHIE spectrograph at Observatoire de Haute Provence, France (PNP.07B.MOUT), and the HARPS spectrograph at ESO La~Silla Observatory (079.C-0127/F). }}

\author{Moutou, C. \inst{1}
\and Bruntt, H. \inst{2}
\and Guillot, T. \inst{3}
\and Shporer, A. \inst{4}
\and Guenther, E. \inst{5}
\and Aigrain, S. \inst{6}
\and Almenara, J.M. \inst{7}
\and Alonso, R. \inst{1}
\and Auvergne, M. \inst{8}
\and Baglin, A. \inst{8}
\and Barbieri, M. \inst{1}
\and Barge, P. \inst{1}
\and Benz, W. \inst{9}
\and Bord\'e, P. \inst{10}
\and Bouchy, F. \inst{11}
\and Deeg, H.J. \inst{7}
\and De la Reza, R. \inst{12}
\and Deleuil, M. \inst{1}
\and Dvorak, R. \inst{13}
\and Erikson, A. \inst{14}
\and Fridlund, M. \inst{15}
\and Gillon, M. \inst{16}
\and Gondoin, P. \inst{15}
\and Hatzes, A. \inst{5}
\and H\'ebrard, G. \inst{11}
\and Jorda, L. \inst{1}
\and Kabath, P. \inst{14}
\and Lammer, H. \inst{17}
\and L\'eger, A. \inst{10}
\and Llebaria, A. \inst{1}
\and Loeillet, B. \inst{1,11}
\and Magain, P. \inst{18}
\and Mayor, M. \inst{16}
\and Mazeh, T. \inst{4}
\and Ollivier, M. \inst{10}
\and P\"atzold, M. \inst{19}
\and Pepe, F. \inst{16}
\and Pont, F. \inst{6}
\and Queloz, D. \inst{16}
\and Rabus, M. \inst{7}
\and Rauer, H. \inst{14,21}
\and Rouan, D. \inst{8}
\and Schneider, J. \inst{20}
\and Udry, S. \inst{16}
\and Wuchterl, G. \inst{5}
}

\offprints{\email{claire.moutou@oamp.fr}}

\institute{LAM, UMR 6110, CNRS/Univ. de Provence, 38, rue F. Joliot-Curie, 13388 Marseille, France
\and
School of Physics A28, University of Sydney, Australia
\and
OCA, CNRS UMR 6202, BP 4229, 06304 Nice Cedex 4, France
\and
Wise Observatory, Tel Aviv University, Tel Aviv 69978, Israel
\and
Th\"uringer Landessternwarte, 07778 Tautenburg, Germany
\and
School of Physics, Univ. of Exeter, Exeter EX4 4QL, UK
\and
Instituto de Astrof\'\i sica de Canarias, E-38205 La Laguna, Spain
\and
LESIA, CNRS UMR 8109, Obs. de Paris, 92195 Meudon, France
\and
Physikalische Institut Univ. Bern, 3012 Bern, Switzerland
\and
IAS, Universit\'e Paris XI, 91405 Orsay, France
\and
IAP, CNRS, Univ. Pierre \& Marie Curie, Paris, France
\and
Observat\'orio Nacional, Rio de Janeiro, RJ, Brazil
\and
IfA University of Vienna, 1180 Vienna, Austria
\and
Institute of Planetary Research, DLR, 12489 Berlin, Germany
\and
RSSD, ESA/ESTEC, 2200 Noordwijk, The Netherlands 
\and
Obs. de Gen\`eve, Univ. de Gen\`eve, 1290 Sauverny, Switzerland
\and
IWF, Austrian Academy of Sciences, 8042 Graz, Austria
\and
IAG, Universit\'e de  Li\`ege, All\'ee du 6 ao\^ut 17, Li\`ege 1, Belgium
\and 
RIU, Universit\"at zu K\"oln, 50931 K\"oln, Germany
\and
LUTH, Obs. de Paris, 5 place J. Janssen, 92195 Meudon, France
\and
ZAA, Technical University Berlin, D-10623 Berlin, Germany
}

\abstract
{}
{The CoRoT satellite has announced its fourth transiting planet (Aigrain et~al. 2008) with space photometry.   
We describe and analyse complementary observations of this system performed to establish the planetary nature of the transiting body and to estimate the fundamental parameters
of the planet and its parent star.}
{We have analysed high precision radial-velocity data, ground-based photometry, and high signal-to-noise ratio spectroscopy.}
{The parent star CoRoT-Exo-4 (2MASS 06484671-0040219) is a late F-type star of mass of 1.16 \Msun\ and radius of 1.17 \Rsun.  The planet has a circular orbit with a period of 9.20205d. The planet radius is 1.19 \Rjup\ and the mass is 0.72 \Mjup. It is a gas-giant planet with a ''normal'' internal structure of mainly H and He. CoRoT-Exo-4b has the second longest period of the known transiting planets.
It is an important discovery since it occupies an empty area in the mass-period diagram of transiting exoplanets.}
{}

\date{Received / Accepted }
\titlerunning{CoRoT-Exo-4b}
\authorrunning{C. Moutou et al.}

\maketitle

\section{Introduction}
\label{sec:intro}

The masses and radii of extrasolar planets are fundamental parameters
for constraining the physics at play in their interiors. A combination of
photometric and Doppler techniques is presently the only way to
indirectly derive these parameters. However, this possibility is
limited to planets that transit in front of their stars, hence a sample of objects that is highly biased towards close-in orbits.

The planet search programme conducted from space by CoRoT
\citep{baglin06} is currently monitoring tens of thousands of stars
and is regularly discovering new transiting planets
\citep{barge08,alonso08,deleuil08b}. CoRoT has the ability to discover planets with smaller
radii and/or longer periods than what is routinely achieved by
ground-based transit surveys. The evolution
of fundamental parameters of extrasolar planets will thus be explored
as a function of the distance to the star and as related characteristics
like irradiation.  This has important implications for our
understanding of these planets. For example, it is a way to test
whether tidal damping (highly dependent on orbital distance),
increased opacities, or alternative theories are the explanation for
the anomalously large sizes of some extrasolar planets
\citep[e.g.][]{guillot06,burrows07,chabrier07}.

With the current detection threshold of the first CoRoT run (45 day duration), 26 transiting events were detected. Two of them turned out to be planets (CoRoT-Exo-1b and 4b, \citet{barge08, aigrain}, 2 are grazing eclipsing binaries, 7 are low-mass binaries, and 10 are background eclipsing binaries. Four events remain unsolved cases, either because they are fast rotators (no possible radial velocity follow-up, the cross-correlation function is too wide or undetected with HARPS) or because they showed no radial-velocity variations. These cases are still in the list of systems to be observed at the next visibility season with higher accuracy. The statistics achieved on this run is compatible with expectations and the simulations on CoRoT detection, with its current threshold and follow-up performances.

The present paper focuses on the fourth planet discovered by
CoRoT. The detection and light curve analysis is described in a
companion paper \citep[ hereafter Paper~IV]{aigrain}. With its 9.20205d
period, the planet CoRoT-Exo-4b has the second longest period known so
far in the family of transiting planets, after HD~17156b
\citep{barbieri07}.  We conducted a series of
complementary observations from the ground to establish the planetary
nature of CoRoT-Exo-4b and to derive the fundamental parameters of
both the star and the planet.

\section{Observations}
\label{sec:obs}
\subsection{Photometric observations}

The photometric measurement was performed on-board CoRoT 
in an aperture that is adapted to each star and is trapeze shaped, 
with a solid angle close to that of a 14 arcsec diameter circle.
It means that a faint eclipsing
binary in the vicinity of the main target can produce a transiting
event diluted enough to mimic a planetary signal \citep{brown03}.
The contamination factor of CoRoT-Exo-4 is estimated from the CoRoT
entry catalogue EXODAT \citep{deleuil08a} to be extremely low,
i.e.\ only 0.3\% of the light comes from neighbouring stars. The first
step in establishing the nature of the transiting body is to check the
photometric behaviour of the stars in the vicinity of the target, and
especially those that fall inside the photometric aperture mask.  This
is easily done with ground-based telescopes. 

CoRoT-Exo-4 was first observed from the Wise
Observatory 18-inch telescope \citep{brosch08} on November 12, 2007.
Only a partial transit was recorded, thus preventing a precise estimate of
the transit depth (of about 1.4\%, which should be compared to the out-of-transit scatter of
0.16\% in 5-min bins), but it is important that no significant variation was
observed in nearby stars.  Another transit event was observed
with the Euler Camera (120-cm Swiss Telescope at La Silla Observatory)
on November 30th, 2007.  Two time series of about 40-min were
collected in the star field, with a reference series taken at a
similar elevation but one night prior to the transit.  The transit
depth was estimated to be $1.2\pm0.15$\%, in agreement with the CoRoT
measurements (1.1\%). Finally, the ingress of another transit was observed at
the Canada-France-Hawaii Telescope with MegaCam on December 10,
2007.  Again, only a partial transit was observed, and the transit
depth was not fully reliable.

The three sets of photometric follow up confirm that
1) the transit observed by CoRoT occurs on the central target at the predicted epochs, 
2) no star encompasses a deep eclipse, or even a short-term variation, in the vicinity of the target, 
3) the transit depth is compatible with the depth measured by CoRoT (i.e.\ no strong dilution by background objects).
A detailed analysis of these data will be given in Deeg et al. (in preparation). 

\begin{figure}
\begin{center}
\epsfig{file=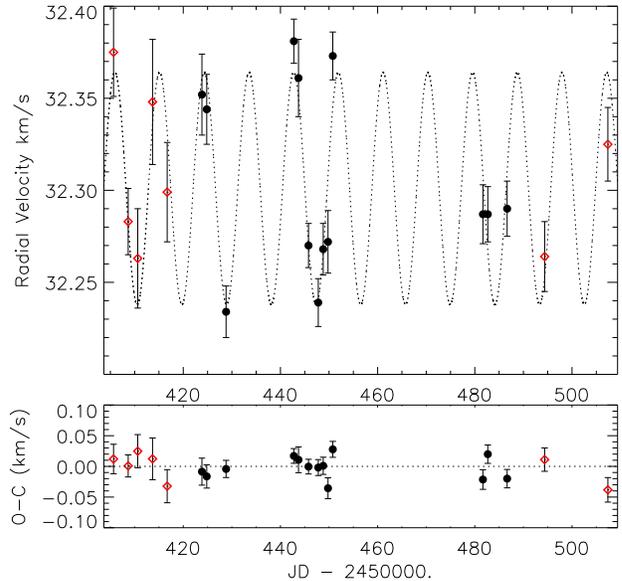,width=0.95\linewidth,angle=0}
\caption{Radial-velocity measurements of CoRoT-Exo-4 as a function of time, showing the best-fit solution and $O-C$ residuals. Black filled circles are HARPS data, and red open diamonds are SOPHIE data.} 
\label{fig:fig2}
\end{center}
\end{figure}

\subsection{Radial velocity observations}

Seven radial velocity measurements of CoRoT-Exo-4 were obtained with SOPHIE on the 1.93m telescope 
at Observatoire de Haute Provence (France), from October 31, 2007 to February 2, 2008. 
These first data points established the planetary nature of the companion, showing a detectable and low-amplitude, 
radial-velocity variation in phase with the CoRoT ephemeris. Thirteen additional measurements 
were also obtained with HARPS (3.6m telescope at La Silla, ESO, Chile), from November 19, 2007 to January 21, 2008. 

Both instruments are cross-dispersed, fiber-fed \'echelle spectrographs on which precise radial-velocity measurements 
 were performed with regular Th-Ar calibration observations and the reduction involves cross-correlation techniques \citep{baranne96,pepe02}. 
With SOPHIE, the high-efficiency mode was used with a resolving power of 40,000, and the sky background 
was simultaneously recorded to correct for contamination \citep{bouchy06}. The radial velocities were 
obtained with a weighted cross correlation of the extracted spectra with a numerical mask corresponding to a G2 star. 

\begin{figure}
\begin{center}
\epsfig{file=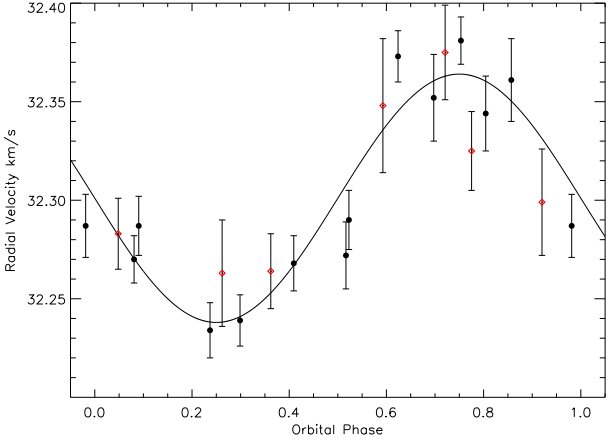,width=0.7\linewidth,angle=90}
\caption{Radial-velocity measurements of CoRoT-Exo-4 with respect to the orbital phase as estimated from the CoRoT ephemeris and the best-fit solution. Black filled circles are HARPS data, and red open diamonds are SOPHIE data.} 
\label{fig:fig3}
\end{center}
\end{figure}

The data are listed in Table~\ref{tab:rv} (available in the on-line version) and shown in 
Fig.~\ref{fig:fig2}. 
The orbital solution was derived using the combined HARPS and SOPHIE datasets. 
 In the fit we included a constant offset between the two sets of measurements.
The ephemeris of the CoRoT light curve was used to constrain the solution. 
The data were best-fitted with a circular Keplerian orbit of semi-amplitude of $63\pm6$\,\ms. 
When using the 13 HARPS data points alone, the residual rms is 17.6\,\ms. 
With the combined SOPHIE and HARPS datasets (20 measurements), the residual rms is 19.4 \ms. 
The offset between SOPHIE and HARPS data is only 5 \ms. 
Figures \ref{fig:fig2} and \ref{fig:fig3} show the data and the best fit as a function of time and orbital phase.
The $O-C$ time series does not show any variations that would indicate a second massive body in the system with a baseline of 100 days (Fig.~\ref{fig:fig2}).

\begin{figure}
\begin{center}
\epsfig{file=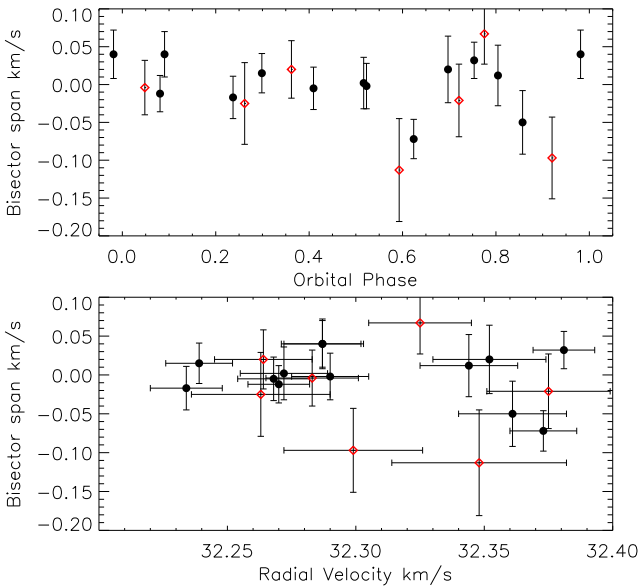,width=0.95\linewidth,angle=0}
\caption{Bisector variations (span of the bisector slope) as a function of orbital phase (top panel) and radial velocity value (bottom panel). Black filled circles are HARPS data, and red open diamonds are SOPHIE data.} 
\label{fig:fig4}
\end{center}
\end{figure}

The bisector analysis is shown in Fig.~\ref{fig:fig4}.  The errors on
the span of the bisector slopes are twice as large as for the
peak position of the cross-correlation function. The behaviour
of the bisector span (BIS) with respect to the radial velocity shows a
dispersed relation with little or no trend. Given that short-term variations are seen in the CoRoT
light curve (see with Fig.~2 in Paper~IV), we tentatively interpret
these variations as possibly stemming from a residual of stellar activity.

\subsection{Determination of spectroscopic parameters}

From radial velocity observations we estimated the projected velocity of the
parent star. The result is $v \sin i = 6.4\pm1$\,\kms, 
based on the width of the cross-correlation functions from SOPHIE and HARPS.
 This agrees with a direct comparison of the observed spectrum with a rotationally broadened synthetic spectrum.
With our determination of the stellar radius $R_\star=1.15$\,\Rsun (see below), this translates into a
rotational period of the star ranging between $P_{\rm rot}=8.5$ and
$10.0$ days. This value for the rotation period is compatible with the
analysis in Paper~IV, and is consistent with a synchronicity of the
stellar rotation and the planet orbit.

We determined the fundamental parameters of the host star
by analysing a HARPS spectrum.
The signal-to-noise (S/N) level in the continuum
is around 60 per pixel in the range 5\,000-6\,000\,\AA\
and it decreases to 40 in the red and blue parts.
The spectrum was carefully normalized
by identifying continuum windows in a synthetic spectrum
calculated with the parameters of the target star.
A spline function was fitted to the continuum windows and
divided into the observed spectrum to obtain the normalized spectrum.
We made sure that the shapes and depths of lines in adjacent
spectral orders were in good agreement. 
In the final spectrum, the orders were merged and 
the overlapping parts added using weights based on the measured S/N.

We analysed the normalized spectrum using
the semi-automatic software package VWA \citep{bruntt04, bruntt08}.
More than 600 mostly non-blended lines were selected
for analysis in the wavelength range 3\,980--6\,830\,\AA.
The VWA uses atmosphere models interpolated in 
the grid by \cite{heiter02} and atomic parameters from
the VALD database \citep{kupka99}.
Abundances were calculated relative to the solar spectrum from
\cite{hinkle00} following the approach by \cite{bruntt08}.
The VWA automatically adjusts the parameters ($T_{\rm eff}$, $\log g$,
microturbulence) of the applied atmospheric model to get consistent results
for all iron lines.
The program iteratively minimizes the correlation between the abundance
and equivalent width found from weak Fe\,{\sc i} lines (equivalent widths
$<80$\,m\AA),
the correlation of abundance and excitation potential of Fe\,{\sc i} lines
(equivalent widths $<140$\,m\AA), and the mean abundance of neutral and
ionized Fe lines must yield the same result. To estimate the uncertainty
on the derived parameters and abundances of individual elements,
we measured the sensitivity of the results when changing
the parameters of the atmospheric model (see \citealt{bruntt08}).

The result is $T_{\rm eff}=6190\pm60$\,K, $\log g=4.41\pm0.05$,
and a microturbulence $\xi_t=0.94\pm0.07$\,km\,s$^{-1}$. The determined spectroscopic parameters do not depend on the adopted value of $v \sin i$.   
The overall metallicity is found as the mean abundance of the elements
with at least 20 lines (Si, Ca, Ti, Cr, Mn, Fe, Ni) yielding
[M/H]\,$=+0.05\pm0.07$.  Details on individual abundances of CoRoT host stars will be given in a later paper.
\section{Estimation of the stellar and planetary parameters}
\label{sec:ana}
\subsection{Stellar mass, radius, and age}

Using evolutionary tracks \citep{siess00,morel07}, the spectroscopic determination 
of $T_{\rm eff}$ and the ratio $M_{s}^{1/3}/R_s$ estimated from the CoRoT photometry (Paper~IV), 
we derived the following estimates for the mass and radius of the star:
$M_s = 1.16_{-0.02}^{+0.03}$\,\Msun\ and $R_s = 1.17_{-0.03}^{+0.01}$\,\Rsun.
 This corresponds to a surface gravity of $\log g= 4.37\pm0.02$, which agrees with the spectroscopic value of $4.41\pm0.05$. 

While the details of the process of how lithium is depleted in
F-stars is not yet fully understood, it is now well-established that
lithium is depleted for stars with a temperature below 6\,200\,K
\citep{steinhauer}.  
 However, lithium depletion in F-stars is expected to be slow, 
and data taken for stars with the same age and spectral type show 
a huge scatter in the equivalent width.
By comparing the equivalent width of the Li\,{\sc i} line at $\lambda$6708
($35.2\pm4.8$ m\AA) of CoRoT-Exo-4b with the data 
for the Hyades \citep{boesgaard,thorburn}, 
the UMa moving group \citep{soderblom}, and NGC~3960 \citep{prisinzano},
an age of about 1$_{-0.3}^{+1.0}$\, Gyr is the most likely estimate for CoRoT-Exo-4b.

\subsection{Planetary mass and radius}

\begin{figure}
\begin{center}
\epsfig{file=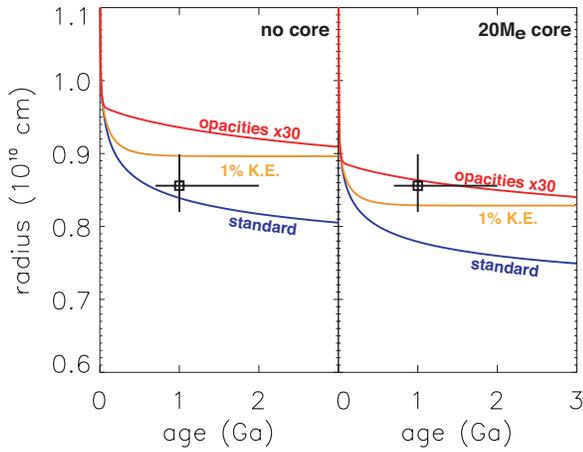,width=0.85\linewidth,angle=0}
\caption{Theoretical models for the contraction of CoRoT-Exo-4b with
  time compared to the observations. The models \citep{guillot06} assume either solar
  composition (left panel) or the presence of a central dense core of
  $20\,\rm M_\oplus$ (right panel). In each case, three possible
  evolution models are shown: (1) a standard model; (2) a model
  assuming that 1\% of the incoming stellar luminosity is transformed
  into kinetic energy and dissipated in the deep interior by tides,
  (3) a model in which interior opacities have been arbitrarily
  increased by a factor 30. 
}
\label{fig:fig5}
\end{center}
\end{figure}

Using the values obtained from the transit fit to the CoRoT light
curve (Paper~IV), the characteristics of the radial-velocity fit, and
the stellar parameters from the previous section, we get a planetary
mass of $M_p = 0.72 \pm 0.08$\,\Mjup\ and a planetary radius $R_p =
1.19_{-0.05}^{+0.06}$\,\Rjup.  The semi-major axis of the orbit is
$a=0.090\pm0.001$\,AU. The mean density of the planet is
$0.525\pm0.15$\,\gcmmm. This new planet is found in the expected
location of the mass-radius diagram \citep{fressin07}. In terms of
composition, Fig.~\ref{fig:fig5} shows that it contains a relatively small
amount of heavy elements (0 to up to $\sim 30\rm\,M_\oplus$, depending
on the model chosen), as expected from the empirical 
correlation between stellar metallicity and mass of heavy elements in
the planet \citep{guillot06,burrows07,guillot08}. The agreement
with different models of planetary evolution implies that we cannot
rule out any specific scenario used to explain anomalously large
planets at this point. However, the discovery of other transiting
planets on long-period orbits will be extremely valuable.

\section{Discussion}
The planet CoRoT-Exo-4b has the second longest period of the
transiting systems known today and is found in a region 
of the mass-period parameter space that was previously empty 
(see Fig.~\ref{fig:fig6} available on-line).   

The mean density of $0.525$\,\gcmmm\ makes CoRoT-Exo-4b
a gas-giant with a ``normal'' internal structure dominated by H and He. It is located in the bulk of Hot Jupiters in the mass-radius diagram. We estimate the age of the parent star to be about 1\,Gyr; thus, the star is likely to be fairly active and its age and rotation rate (Paper IV) are compatible.

New discoveries made with CoRoT will likely populate the mass-period diagram in
the period range 5--50 days and allow further investigations of
the importance of star-to-planet distance on the internal structure of
giant planets.  This will complement the output from current
radial-velocity surveys, allow us to better understand the mechanisms
that govern the contraction of these planets, and provide constraints
for planet formation models.

\begin{table}
\begin{minipage}[t]{\columnwidth}
\begin{center}
\caption{The orbit parameters and the fundamental parameters of CoRoT-Exo-4b and its host star.}  
\renewcommand{\footnoterule}{}  
\begin{tabular}{llll}
\hline\hline
$V_0$ [\kms]        & $32.301   \pm 0.005     $&  $K$ [\ms]           & $63      \pm 6         $ \\
$e$                 & $0.0       \pm 0.1       $ & $a$ [AU]            & $0.090  \pm 0.001    $ \\ 
$P$ [d] & 9.20205 $\pm$ 0.00037& & \\ 
$M_p$ [\Mjup]       & $0.72    \pm 0.08       $ &  $R_p$ [\Rjup]       & $1.190_{-0.05}^{+0.06}$ \\
$\rho_p$ [g/cm$^3$] & $0.525   \pm 0.15      $ & $T_{\rm eq}$ [K]    & $1,074    \pm 19       $ \\
\hline 					      
$M_s$ [\Msun]       & $1.16_{-0.02}^{+0.03}  $ & $R_s$ [\Rsun]       & $1.17_{-0.03}^{+0.01}  $ \\
$v\sin i$ [km/s]    & $6.4     \pm 1.0       $ & $T_{\rm eff}$ [K]   & $6190    \pm 60        $ \\
$\log g$            & $4.41    \pm 0.05      $ & Age  [Gy] &  $1_{-0.3}^{1.0}$\\ 
\hline \hline
\label{tab:param}
\end{tabular}
\end{center}
\end{minipage}
\end{table}

\onlfig{5}{
\onecolumn
\begin{figure}
\begin{center}
\epsfig{file=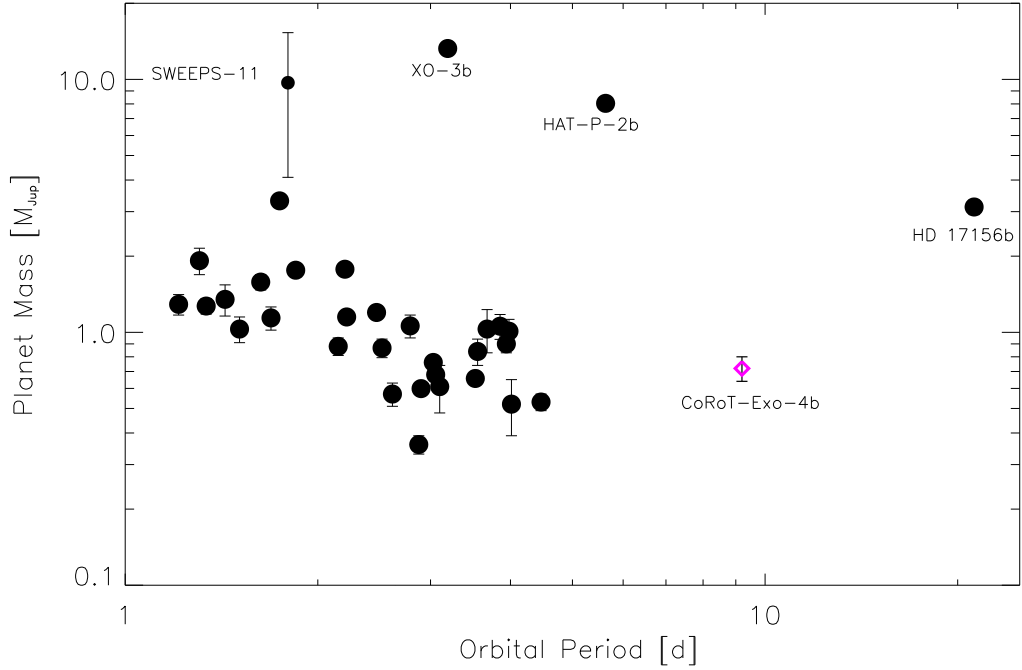,width=0.8\linewidth,angle=0}
\caption{Mass-period diagram of the 35 known transiting planets showing the location of CoRoT-Exo-4b (diamond).} 
\label{fig:fig6}
\end{center}
\end{figure}
}

\onltab{2}{
\begin{table}
\begin{center}
\caption[]{Radial-velocity measurements of CoRoT-Exo-4 with SOPHIE (lines 1 to 7) and with HARPS (lines 8 to 20). "BIS" is the slope of the bisector span, measured in the cross-correlation function.  }  
\begin{tabular}{lllll}
\hline\hline
\small
BJD &  RV & Uncertainty & BIS & err(BIS)\\
 - 2400000 &      [km~s$^{-1}$] & [km~s$^{-1}$]&[km~s$^{-1}$]&[km~s$^{-1}$] \\
\hline \hline
54405.6545 &  32.370&0.024&  $ -0.021$ & 0.048 \\
54408.6667 &  32.278&0.018&  $ -0.004$ & 0.036 \\
54410.6344 &  32.258&0.027&  $ -0.025$ & 0.054 \\
54413.6784 &  32.343&0.034&  $ -0.113$ & 0.068 \\
54416.6908 &  32.294&0.027&  $ -0.097$ & 0.054 \\
54494.3741 &  32.259&0.019&  $ +0.020$ & 0.036 \\
54507.3778 &  32.320&0.020&  $ +0.067$ & 0.040 \\
\hline
54423.8430&   32.352&0.022&   $ +0.020$ & 0.044 \\
54424.8273&   32.344&0.019&   $ +0.012$ & 0.040 \\
54428.8091&   32.234&0.014&   $ -0.017$ & 0.028 \\
54442.7618&   32.381&0.012&   $ +0.032$ & 0.024 \\
54443.7161&   32.361&0.021&   $ -0.050$ & 0.042 \\
54445.7748&   32.270&0.012&   $ -0.012$ & 0.024 \\
54447.7813&   32.239&0.013&   $ +0.015$ & 0.026 \\
54448.7990&   32.268&0.014&   $ -0.005$ & 0.028 \\
54449.7871&   32.272&0.017&   $ +0.002$ & 0.034 \\
54450.7729&   32.373&0.013&   $ -0.072$ & 0.026 \\
54481.6668&   32.287&0.016&   $ +0.040$ & 0.032 \\
54482.6719&   32.287&0.015&   $ +0.040$ & 0.030 \\
54486.6509&   32.290&0.015&   $ -0.002$ & 0.030 \\
\hline\hline
\label{tab:rv}
\end{tabular}
\end{center}
\end{table}
}
\begin{acknowledgements}
We thank the OHP and ESO staff for help provided during the SOPHIE/HARPS observations. 
The German CoRoT team acknowledges support through DLR grants 50OW0204, 50OW0603 and 50QP0701. 
HJD and JMA acknowledge support by grants ESP2004-03855-C03-03, and ESP2007-65480-C02-02 of the Spanish Education \& Science ministry.
\end{acknowledgements}
\bibliographystyle{aa}
\bibliography{references}

\end{document}